\documentclass{article}

\usepackage{PRIMEarxiv}

\usepackage{amsmath}
\usepackage{makecell}
\usepackage{bm}

\usepackage[utf8]{inputenc} 
\usepackage[T1]{fontenc}    
\usepackage{hyperref}       
\usepackage{url}            
\usepackage{booktabs}       
\usepackage{amsfonts}       
\usepackage{nicefrac}       
\usepackage{microtype}      
\usepackage{lipsum}
\usepackage{fancyhdr}       
\usepackage{graphicx}       
\graphicspath{{media/}}     

\title{Exit Spillovers of Foreign-invested Enterprises in Shenzhen's Electronics Manufacturing Industry
\thanks{I thank Prof. Samantha A. Vortherms for her invaluable guidance and support with this paper.} 
}

\author{
  Hanqiao Zhang \\
  University of California Irvine \\
  Irvine, CA\\
  \texttt{hanqiaoz@uci.edu} \\
}

\begin{document}
\maketitle

\begin{abstract}
Neighborhood characteristics have been broadly studied with different firm behaviors, e.g. birth, entry, expansion, and survival, except for firm exit. Using a novel dataset of foreign-invested enterprises operating in Shenzhen's electronics manufacturing industry from 2017 to 2021, I investigate the spillover effects of firm exits on other firms in the vicinity, from both the industry group and the industry class level. Significant neighborhood effects are identified for the industry group level, but not the industry class level.
\end{abstract}

\keywords{Neighborhood Effects \and Firm Exit \and Industry Agglomeration \and Spatial Econometrics}

\section{Introduction}

Spatial dependence on various firm behaviors, such as birth, entry, survival, and expansion, is widely studied in the context of the influences of local characteristics, industry agglomeration, and specialization. However, few papers have associated it with firm exit. Empirical evidence shows significant geographical patterns and neighborhood effects for firm exit. \cite{arcuri2019spatial} investigates firm exits in France and discovers that places with high exit rates are more likely to be surrounded by similar ones. \cite{sarmiento2007spatially} concludes the spatial binary lagged dependent variable is the most powerful in explaining firm exit in the U.S. baking industry.

The contribution of this chapter is three-fold. First of all, using a new dataset of foreign-invested enterprises in Shenzhen's electronics manufacturing industry, the paper adds evidence to the existence of spatial dependence on firms' exit behaviors. Secondly, it depicts a picture of divestment in Shenzhen's manufacturing industry across time in the context of foreign-invested enterprises' exits. I analyze neighborhood effects and important factors to firm exits under different hierarchies of industry classification categories. Thirdly, it looked into how neighborhood effects of firm exit behaviors may be impacted by large external shocks. During the past 5 years, at least two external shocks impacted all foreign-invested enterprises that were operating in mainland China: raised tariffs and political risks brought by the 2018 U.S.-China trade war, and rapid changes in the business environment induced by COVID-19. The shocks not only affect firms' probability to exit directly, but they may also generate non-trivial neighborhood effects. For example, tariffs imposed on a specific industry result in the leaving of some firms from highly specialized production geography, and these firm exits may increase or decrease the likelihood of other firms' leaving due to the loss of benefits from specialization, disruption of the supply chain, or the growth of market share freed by former competitors.

The second section reviews the literature that is closely related to the spillover effects of foreign-invested enterprise exits. The third section theorizes the spillover effects on firm exits. The fourth section introduces the novel foreign-invested enterprises in China dataset. The fifth section describes the spatial lagged probit model and the applied GMM estimator. The sixth section illustrates the empirical results, and the last section concludes.

\section{Literature Review}

Firm exit is one of the most discussed topics since \cite{baldwin1991firm}, and its determinants have been studied in a large body of literature, see a systematic review in \cite{cefis2022understanding}. Although local context has been studied extensively for firm birth \cite{cala2016regional,audretsch2015regional,lee2013regional,levratto2014does}, firm entry \cite{cheratian2021spatial}, firm survival \cite{huiban2011spatial,craioveanu2016impact}, and firm growth \cite{levratto2016does}, it has not been related to the firms' exit behaviors until recent years \cite{weterings2015spatial,ferragina2015agglomeration}.

Two strands of related literature are industrial agglomeration and specialization. The former refers to the phenomenon that firms tend to cluster geographically \cite{audretsch1996r,porter1998clusters}, and the benefits may outweigh the disadvantages brought by higher industrial densities, such as less shared resources and fiercer competition. Micro-foundations are offered by \cite{duranton2004micro} based on Marshall's trinity \cite{marshall2009principles}: matching, sharing, and learning. Industrial clustering could produce better outcomes in matching employer and employee in terms of both the matching quality and probability \cite{rosenthal2001determinants}. Firms may share a wide variety of input suppliers, expensive facilities, and the gains of individual specialization \cite{baumgardner1988division}. The latter emphasizes the concentration of enterprises in a particular industry or sector in a given region, which could facilitate innovation \cite{duranton2001nursery}, knowledge spillovers \cite{jovanovic1989growth}, infrastructure in the region, and accumulation. Empirical results also provide supporting evidence. \cite{cainelli2014spatial} found that specialization decreased firm exit rates in the short run, especially for the low-tech industry. In the case of \cite{power2019effect}, specialization reduces exits at the firm level but not the regional level. In the context of China, \cite{fan2003industrial} substantiates positive relationships between spatial agglomeration and productivity in various Chinese manufacturing sectors. To my best knowledge, the only paper that directly estimates spatial correlations of firm exit is \cite{arcuri2019spatial}, in which both the dependent variable, exit rate, and control variables are at the aggregated French department level. They identify significant positive spatial autocorrelation of firm exits. Compared to the previous study, this paper models the spatial dependence of enterprises' exit behaviors using firm-level observation and control variables.

Focusing on the closure of foreign-invested enterprises, this paper also fits into the fast-growing foreign divestment literature. Compared to foreign direct investment, comparatively little attention has been paid to this area due to data limitations. Observations of enterprises' divestment not only require panel data \cite{lee2010divestiture}, but also need the enterprises to willingly share divestment decisions that may show business failures \cite{benito1997divestment}. Some theories that explain foreign subsidiary divestment highlight the importance of local market conditions and interconnections among enterprises. From the resource-based point of view, \cite{barney1991firm} identifies four empirical indicators that determine a foreign subsidiary's potential to generate competitive advantage and thus could protect it from divestment: value, rareness, imitability, and substitutability. Eclectic paradigm, developed by \cite{dunning1980toward}, considers three factors – ownership, location, and internalization – that influence the decision-making of multinational enterprises when engaging in foreign divestment decisions. When an enterprise divests from a region, it may signal a decline in the value of its advantages, prompting other enterprises with similar resources to reconsider their presence in the region. Unfortunately, few empirical papers have yet been found to focus on the relationship among enterprises' divestment decisions in the same region, or how remaining enterprises may act in response to the divestment of other enterprises in the vicinity.

\section{Theorizing Neighborhood Effects on Enterprise Exits}

Exit decisions of enterprises in the neighborhood may not be made independently. Research in industrial agglomeration and specialization shows that enterprises tend to locate near others that produce either homogeneous or similar products of different magnitudes due to the benefit brought by spillover effects or economies of scale. If regional enterprise density declines due to the exit of neighbors, an enterprise may no longer enjoy the positive externalities, such as knowledge spillovers, access to suppliers and customers, larger labor pool, etc., and decide to exit itself.

Neighborhood effects on exit decisions could vary by the business relationships among the enterprises in the same industry. On one hand, the exits of direct competitors could create a void that allows the remaining firms to capture a larger market share. If more competitors are leaving, the overall competitiveness of the regional market may even be impacted. It becomes easier for the remaining enterprises to collude and increase their market power, making them less likely to exit the market. On the other hand, the exit of an enterprise in a neighborhood could disrupt the local supply chains that other parties in the area have been relying on. For example, if the exiting enterprise was a supplier or customer of other enterprises in the area, the remaining enterprise may also consider leaving because it struggles to find alternative suppliers or customers, and thus may have higher costs and lower profitability. The business relationships among enterprises in the same industry may not be identical according to the definition of the industry, or its aggregate level in the data. The lower the aggregate level, the more subdivided industry the enterprises may operate within, thus having a stronger intensity of competition. Higher industry aggregate level may bring in more indirect competitors who produce non-homogeneous goods but satisfy the same general demand, or even support enterprises that operate in the upstream or downstream subdivided industry. In this case, the competition intensity within the same industry would be lower, and connections among enterprises would be more complicated.

In the presence of neighborhood effects, the U.S.-China trade war in 2018 and 2019 could be a significant disruptor to regional industrial clusters beside the tariff impositions. As some enterprises exit the market due to higher tariffs, it would be more difficult for the remaining firms to maintain economies of scale, and keep lower costs of sourcing raw materials or components. A downward spiral of economic activity in the surrounding area may occur, e.g. demand for related services like logistics and transport is reduced. This could further weaken the economic viability of the staying firms, contributing to their probability of exiting and creating a domino effect.

Apart from the spillover effects generated by nearby enterprises, the exit decision could be affected by a series of enterprise-level variables as well. Firstly, more productive and efficient enterprises are less likely to exit the market because they could better compete with other enterprises in the market by producing at a lower cost, offering higher quality products or services, and investing more in research and development. Scholars found that higher firm productivity, both in technical and labor efficiency, is related to higher survival time and a lower rate of market exits \cite{muzi2022productivity}, \cite{aga2017market}. Secondly, the way firms are structured and governed may play an important role in their exit decisions. A sole proprietorship or partnership may be more likely to exit the market compared to a corporation because the liabilities of the former are tied to the personal assets of the owner, whereas the liabilities of the latter are limited to the assets of the corporation. It could make it riskier for small enterprises to stay in the market if they face financial difficulties or other challenges. There is also some relevant empirical evidence. \cite{cotei2018m} finds firm's legal structure, such as operating as a corporation or sole proprietorship, affects its acquisition outcome through innovation and employment growth. \cite{goktan2018corporate} argues that more than many economic factors, some corporate governance features, such as the size of independent boards, and restrictions on shareholder governance, are more important in determining if a company would exit by M\&A, going private, or going bankrupt. Thirdly, enterprises' ages and sizes could be related to their probability to exit. On one hand, younger and smaller enterprises may be more likely to exit the market, because they are typically less established and have less experience. They may face greater challenges in accessing resources, such as capital and customers, or be more vulnerable to changes in the market or competition. On the other hand, older enterprises may exit if they are less flexible and adaptable compared to younger ones, making them difficult to respond to changes or adapt to new technologies. Research also shows that foreign enterprises become less attached to local markets as they age, grow "footloose" and have higher market exit rates, see \cite{mata2012foreignness}, \cite{coucke2008offshoring}.

\section{Foreign Electronics Manufacturer in Shenzhen}

The dataset is constructed with two sources: the foreign-invested enterprises in China (FIEC) database provided by \cite{vortherms2021political}, and Qichacha, a website that delivers business data on China-based private and public companies. The FIEC dataset covers officially-registered multinational corporations, from 2014 to 2019, that operate in Mainland China and have at least one foreign investor. Reports are collected annually from the website of the Chinese Ministry of Commerce, which requires every foreign-invested enterprise to register its information before June. This includes name in both English and Chinese, date of registration, date of establishment, industry, business practice, firm type, address, registered capital, realized capital, area of registration, investors, and annual reports. Qichacha is an information query platform that collects and collates public information from government websites, such as the National Enterprise Credit Information Publicity System, China Executive Information Disclosure network, etc. I combine the FIEC dataset with the lists of enterprises invested by foreign investors and investors from Hong Kong, Macao, and Taiwan exported from Qichacha as of January 2023. This includes filling in missing values in the FIEC dataset and updating outdated registration information, such as business scope, address, and annual reports. The industry descriptions in the FIEC database also enable the industry classification of foreign-invested enterprises on four different aggregate levels based on the Chinese standard for industrial division\footnote{This refers to the "GB/T 4754—2017 Industrial Classification for National Economic Activities".}. The standard classifies enterprises based on the productive activities they engage, thus enterprises that are grouped together primarily perform the same type of production. The following figure is a minimal example of the four industry aggregate levels in the context of manufacturing.

\begin{figure}[!htbp]
  \centering
  \includegraphics[width=\textwidth]{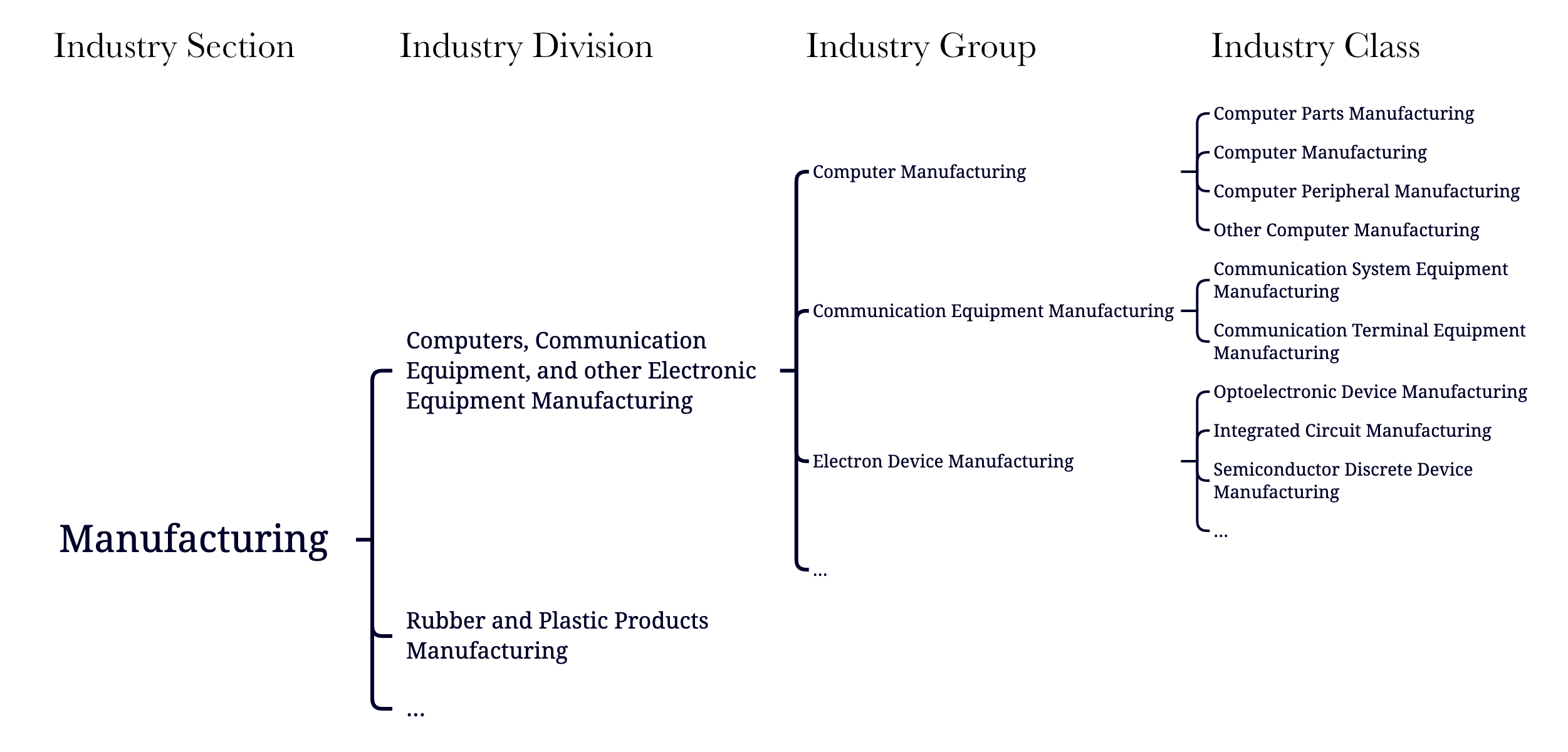}
  \caption{Four Levels of Classification Categories for the Manufacturing Industry-section}
\end{figure}

Firms are more homogeneous in terms of the productive activities that they engage in if the classification category that they belong to is more narrow or specialized. For example, if both firms from the same city specialize in computer parts manufacturing, they may cater to the same group of customers, such as original equipment manufacturers, computer repair and upgrade shops, etc. However, firms are more diversified if we move to broader levels of classification categories. At the industry group level, firms could still operate in the same industry class, or they could be in the supporting industries of another: computer parts manufacturers may be the local suppliers of computer manufacturers.

In the present study, we focus on the enterprises that operate in the electronics manufacturing industry in Shenzhen, 2017, for a few reasons. Firstly, as a major hub for electronics manufacturing in Guangdong province, Shenzhen has a high density of industrial parks and special economic zones. This unique landscape facilitates a significant concentration of foreign-invested enterprises in the region. Secondly, the electronics manufacturing industry has a complex and extensive supply chain, with multiple levels of suppliers and subcontractors. This makes the industry more susceptible to spillover effects, as the exit of a key foreign-invested enterprise could disrupt the supply chain and create uncertainties for other enterprises. Thirdly, in the background of the U.S.-China trade war starting in 2018, the electronics manufacturing industry faces increasing pressure due to geopolitical tensions and trade disputes, and the spillover effect may bring collateral damage that leads to an additional number of firm exits. These contexts make the industry more relevant to study the neighborhood effects of foreign-invested enterprises' exits.

Prior to 2019, the Chinese Ministry of Commerce website provides a full list of foreign-invested enterprises that are operating in mainland China in the current year, so the exit time of each enterprise is marked as the year at which it no longer shows up in the list. Starting in 2020, the website realizes a technology upgrade and displays the search result only for the typed-in enterprise. Therefore, the exit year of each enterprise that remains active through 2019 is marked as the last year that it has submitted its annual report that could be verified on Qichacha. Depending on the identity of foreign investors, the exit behavior could be the subsidiary closure of a foreign multinational corporation, or the abandonment of the Chinese market by a wealthy foreign individual.

For enterprises that operate as early as 2017 in Shenzhen, I report the numbers and percentages of enterprise exit each year from the computers, communication, and other electronic equipment manufacturing division, the broader manufacturing section, and all industry sections, in the following table. For a broad overview of enterprises' exits in all Chinese first-tier cities from 2014 to 2021, see \ref{exits_overview}.

\begin{table}[!htbp] \centering
  \caption{Foreign-invested Enterprises' Yearly Exits in Shenzhen, 2017-2021}
  \setlength\tabcolsep{2pt}
  \begin{tabular}{lcccccc}
    \\[-1.8ex]\hline \hline \\[-1.8ex]
    & \multicolumn{2}{c}{\makecell[c]{Computers, Communication \& Other\\Electronic Equipment Manufacturing\\(Division)}} & \multicolumn{2}{c}{\makecell[c]{Manufacturing\\(Section)}} & \multicolumn{2}{c}{All Industry Sections} \\
    & \multicolumn{1}{c}{Enterprises} & \multicolumn{1}{c}{\makecell[c]{Exits\\(\%)}} & \multicolumn{1}{c}{Enterprises} & \multicolumn{1}{c}{\makecell[c]{Exits\\(\%)}} & \multicolumn{1}{c}{Enterprises} & \multicolumn{1}{c}{\makecell[c]{Exits\\(\%)}} \\
    \hline \\ [-1.8ex]
    2017-2018 & 1,458 & 92 & 8,097 & 441 & 46,889 & 1,727 \\
    & & (5.94\%) & & (5.17\%) & & (3.55\%)  \\
    2018-2019 & 1,300 & 158 & 7,301 & 796 & 42,270 & 4,619  \\
    & & (10.84\%) & & (9.83\%) & & (9.85\%)  \\
    2019-2020 & 1,170 & 130 & 6,434 & 867 & 38,109 & 4,161  \\
    & & (10.00\%) & & (11.88\%) & & (9.84\%) \\
    2020-2021 & 1,116 & 54 & 6,109 & 325 & 36,116 & 1,993  \\
    & & (4.62\%) & & (5.05\%) & & (5.23\%)  \\
    \hline \hline \\[-1.8ex]
  \end{tabular}
\end{table}

Following the outbreak of the trade war in 2018 and the pandemic in 2019, the exit rates in these two years are both at the highest level. The electronics manufacturing industry and the broader manufacturing industry have higher exit rates than the overall industries in general.

Summary statistics of the firm-level characteristics are shown in the table. Distributions of capital size and aggregate investment of the enterprises have a mass below 1,000 and long right tails. $81.16\%$ of the enterprises are foreign-owned, $14.19\%$ are joint-venture companies, and $4.65\%$ have other types, such as contractual joint venture, share-holding, and partnership. For joint ventures, the median percentages of the foreign contribution of both the registered capital and realized capital in joint ventures are about $45\%$. As for places of origin, most of the foreign-invested enterprises are registered in Asia. More than half of the enterprises ($64.52\%$) are registered in Hong Kong, followed by Taiwan ($5.35\%$), British Virgin Islands ($5.03\%$), Samoa ($3.81\%$), and the United States ($2.32\%$). Following \cite{vortherms2021political}, I also include rough indicators of tariff exposure during the 2018 trade war and enterprises' importer/exporter status. The former marks enterprises that operate within tariffed industry class, using tariff data from \cite{bown2019us}, and the latter filters if enterprises mention the keywords "export" or "import" in their business scopes. More than $94\%$ of the enterprises are exposed to tariffs according to the industry class, and at least $51\%$ engages in import and export trade business.

\begin{table}[!htbp] \centering
  \caption{Summary Statistics of Categorical Variables}
  \begin{tabular}{lc}
    \\[-1.8ex]\hline \hline \\[-1.8ex]
    & \multicolumn{1}{c}{Shenzhen} \\
    \hline \\ [-1.8ex]
    Foreign-owned & 81.16\%     \\
    Joint-venture & 14.19\%     \\
    U.S. Registered & 2.32\%    \\
    Tariffed Industry & 94.39\% \\
    Importer/Exporter & 51.55\% \\
    \hline \hline \\[-1.8ex]
  \end{tabular}
\end{table}

\begin{table}[!htbp] \centering
  \caption{Summary Statistics of Numerical Variables}
  \begin{tabular}{lccc}
    \\[-1.8ex]\hline \hline \\[-1.8ex]
    & \multicolumn{1}{c}{Mean} & \multicolumn{1}{c}{Median} & \multicolumn{1}{c}{St.Dev.} \\
    \hline \\ [-1.8ex]
    Aggregate Investment & 13,302.90 & 760.00 & 172,804.50 \\
    Registered Capital & 6348.06 & 547.00 & 64785.70 \\
    Foreign Contribution(\%) for Joint-venture & 47.83 & 45.00 & 0.24 \\
    Realized Capital & 1000.00 & 103.10 & 7748.32 \\
    Foreign Contribution(\%) for Joint-venture & 47.05 & 45.71 & 0.25 \\
    \hline \hline \\[-1.8ex]
    \textit{Note:}  & \multicolumn{3}{r}{All currencies are in \$10,000} \\
  \end{tabular}
\end{table}

\section{Spatial Lagged Probit Model}

To estimate the neighborhood effects on enterprise exits, I apply the classic spatial lagged probit model in spatial econometrics literature \cite{anselin1988spatial, anselin2013advances, lesage2009introduction}. There are $n$ foreign-invested enterprises in Shenzhen's electronic manufacturing industry that remain active in a given year. The observed dichotomous exit decision, $Y_i$, for enterprise $i=1,2,\cdots,n$, depends on the value of a latent continuous variable, interpreted as its propensity to exit, $Y_i^*$:
\begin{equation}
  \nonumber
  Y_i = \left \{ \
  \begin{aligned}
    1, & \quad Y_i^* \geq 0 \\
    0, & \quad Y_i^* < 0 \\
  \end{aligned} \right.
\end{equation}

Enterprises' latent propensity to exit in vector form, $\bm{Y^*}$, is assumed to be the linear combination of itself and the matrix of firm-level control variables, $\bm{X}$:
\begin{equation}
  \nonumber
  \begin{aligned}
    \bm{Y^*} \  & = \  \bm{ \rho W Y^* } + \bm{ X \beta } + \bm{ \epsilon } \\
    \bm{ \epsilon } \ & \sim \  \text{MVN} \left( \ \bm{0}, \  \sigma_{\bm{\epsilon}}^2 \bm{I} \  \right) \\
  \end{aligned}
\end{equation}
where $\bm{\epsilon}$ is assumed to follow a multivariate normal distribution, and $\sigma_{\bm{\epsilon}}^2$ is the variance of the error term. $\bm{X}$ includes enterprises' years of operation, the size of registered capital, the percentage of registered capital contributed by foreign investors, regions of registration, legal form, the tariff indicator, and the importer/exporter indicator. $\bm{W}$ is an exogenously specified weight matrix with $0$s on the diagonal. I define elements in $\bm{W}=\left[W_{ij}\right]$ as the inverse geographical distance of enterprise $i$ and $j$:
\begin{equation}
  \nonumber
  W_{ij} = \left \{ \
  \begin{aligned}
    \frac{1}{d_{ij}}, & \quad \text{if $i$ and $j$ belong to the same industry} \\
    0, & \quad \text{if $i$ and $j$ are from different industries} \\
  \end{aligned} \right.
\end{equation}
where $d_{ij}$ is the great-circle distance computed from the two enterprises' longitudes and latitudes. $\bm{W}$ is then row-normalized as $W_{ij}/\sum_j W_{ij}$, so that the enterprise's propensity to exit is a weighted average of neighboring enterprises' propensities to exit, excluding itself. Geographically closer enterprises generate stronger effects by assumption. Four models are estimated for a given year using four $\bm{W}$ that allow for the correlation of enterprises' propensities to exit within four aggregate levels of the industry respectively.

It was shown that spatial dependence could only be introduced through the latent variable, e.g. models like $\bm{Y^*}=\bm{ \rho W Y }+\bm{ X \beta }+\bm{\epsilon}$, or $\bm{Y}=\bm{ \rho W Y }+\bm{ X \beta }+\bm{\epsilon}$ are not algebraically consistent, see \cite{beron2004probit, klier2008clustering}. In our case, the spatial lagged model assumes the propensity of each firm to exit the market depends on other nearby firms' propensities to exit, instead of whether other nearby firms have actually left. The application is appropriate because firms' market exit decisions, in the language of \cite{klier2008clustering}, are "forward-looking in nature". For example, when an enterprise decides whether to exit a specialized town, it may expect that neighboring enterprises are reluctant to leave and give up the benefits brought by agglomeration. The enterprise anticipates the low values of $\bm{Y^*}$ could hold the cluster or even attract more firms, which further reduces fixed costs or generates greater competitive advantages.

The reduced-form equation of the spatial lagged probit model could be derived:
\begin{equation}
  \nonumber
  \bm{Y_i^*} \  = \  \bm{X^*\beta} + \bm{u}
\end{equation}
where $\bm{X^*}=\left(\bm{I}-\rho\bm{W}\right)^{-1} \bm{X}$, $\bm{u} \sim \text{MVN}\left(0,\bm{\Sigma}\right)$, and $\bm{\Sigma}=\left[\left(\bm{I}-\rho\bm{W}\right)'\left(\bm{I}-\rho\bm{W}\right)\right]^{-1}$.

The model leads to inconsistent and inefficient estimates due to heteroskedastic errors. Many efforts were made to overcome the problems induced by spatial dependence from the aspects of both theoretical and empirical \cite{fleming2004techniques}, and a few estimators were proposed. From the Bayesian's perspective, \cite{mcmillen1992probit} proposes an EM algorithm that replaces the latent $\bm{Y^*}$ with expected values in the E-step, then estimates model parameters with maximum likelihood in the M-step. \cite{lesage2000bayesian} suggests a Gibbs sampler that produces random draws of $y_i^*$ from a multivariate truncated normal distribution conditional on all other model parameters. \cite{beron2004probit} comes up with a recursive importance sampling (RIS) algorithm that directly evaluates the probit likelihood function. From the frequentist's perspective, \cite{pinkse1998contracting} proposes a generalized method of moments (GMM) estimator based on the spatial error probit model, and it was later linearized around zero interdependence in \cite{klier2008clustering}.

The RIS and Bayesian strategies are shown to be able to provide accurate estimates for spatial lagged probit models. The GMM estimators, being instrumented-approximation methods, work well only when the samples are large and the spatial dependence is not strong. However, they are computationally much more efficient with running times orders of magnitude shorter \cite{calabrese2014estimators}. For this reason, I apply the linearized GMM estimator developed in \cite{klier2008clustering} to the data, because it is the only feasible one in estimation time to run one model for each year at each classification category. Spatial dependence parameters from two randomly chosen models are estimated by the Gibbs sampler, and differences between the results and the one generated by the linearized GMM estimator are at the second digit. Nevertheless, this is still considered a limitation of the estimation strategy of this paper, and more models should be run to ensure the GMM estimates are valid.

I then illustrate the GMM estimator for the spatial lagged probit model in detail. Its log-likelihood function could be easily written from the reduced-form equation as:
\begin{equation}
  \nonumber
  \begin{aligned}
    l\left(\bm{\beta},\rho|\bm{X,W,Y^*}\right) \ & = \ \sum_{i=1}^n \left\{ \  y_i \cdot ln P\left(y_i=1|\bm{x_i},W_{ij},y_j^*\right) + \left(1-y_i\right) \cdot ln \left[ 1-P\left(y_i=1|\bm{x_i},W_{ij},y_j^*\right) \right] \ \right\} \\
    P\left(y_i=1|\bm{x_i},W_{ij},y_j^*\right) \ & = \  P\left(y_i^*\geq0|\bm{x_i},W_{ij},y_j^*\right) = P\left(\bm{x_i^{*\prime}\beta}+u_i\geq0|\bm{x_i},W_{ij},y_j^*\right) = \Phi\left( \frac{\bm{x_i^{*\prime}\beta}}{\sigma_i} \right) \\
  \end{aligned}
\end{equation}
where $\Phi(\cdot)$ is the CDF of standard normal distribution, $\sigma_i$ is the $i$th standard deviation according to $\bm{\Sigma}$. \cite{pinkse1998contracting} derived the sample moment condition that accounts for the heteroscedastic errors:
\begin{equation}
  \nonumber
  m\left(\beta,\rho\right) \ = \ \frac{1}{n}\sum_{i=1}^n \bm{z_i'}\left\{ \frac{\left[y_i-\Phi\left(\frac{\bm{x_i^{*\prime}\beta}}{\sigma_i}\right)\right] \phi\left(\frac{\bm{x_i^{*\prime}\beta}}{\sigma_i}\right)}{\Phi\left(\frac{\bm{x_i^{*\prime}\beta}}{\sigma_i}\right) \left[1-\Phi\left(\frac{\bm{x_i^{*\prime}\beta}}{\sigma_i}\right)\right] } \right\}
\end{equation}
where $\bm{z_i}$ is the $i$th row of $\bm{Z}$, the matrix of instruments that are composed of the control variables $\bm{X}$. The parameters of interest satisfy the moment condition $m(\beta,\rho)=0$. When the number of moment conditions exceeds the number of unknown parameters, the model could be estimated by:
\begin{equation}
  \nonumber
  \text{argmin}_{\beta,\rho \in \Theta} \quad  \bm{m}\left(\beta,\rho\right)'
  \bm{M} \bm{m}\left(\beta,\rho\right)
\end{equation}
where $\Theta$ is the parameter space, $\bm{M}$ is a positive-definite matrix that assigns weights to different moment conditions. This equation does not have an analytical solution and needs to be solved with non-linear optimization algorithms. The computation could be burdensome because each iteration involves the inverse of a $n$ by $n$ matrix, $\left(\bm{I}-\rho\bm{W}\right)^{-1}$, while evaluating any candidate value of $\rho$. If $M$ is specified as $\left(\bm{Z}'\bm{Z}\right)^{-1}$, the estimator becomes non-linear two-stage least squares (\cite{amemiya1975nonlinear}, \cite{amemiya1974nonlinear}) with the objective function:
\begin{equation}
  \nonumber
  \text{argmin}_{\beta,\rho \in \Theta} \quad  \hat{\bm{e}}\left(\beta,\rho\right)' \bm{Z}
  \left(\bm{Z}'\bm{Z}\right)^{-1} \bm{Z}' \hat{\bm{e}}\left(\beta,\rho\right)
\end{equation}
where $\hat{\bm{e}}\left(\beta,\rho\right)$ is the generalized probit residuals inside the curly brackets of $m\left(\beta,\rho\right)$. To estimate the parameters, we first assume the initial value $\bm{\Gamma}_0=\left(\rho,\bm{\beta}\right)'$, then compute $\hat{\bm{e}}_0\left(\beta,\rho\right)$ and gradient terms $\bm{G}=\frac{\partial \bm{P}}{\partial \bm{\Gamma}}$. Regress $\bm{G}$ on $\bm{Z}$ to obtain the predicted value $\hat{\bm{G}}$, and the new estimates $\bm{\Gamma}_1=\bm{\Gamma}_0+\left( \hat{\bm{G}}'\hat{\bm{G}} \right)^{-1} \hat{\bm{G}}' \hat{\bm{e}}_0\left(\beta,\rho\right)$. This process is iterated until convergence. This GMM estimator is computationally challenging because the gradient term,  $\frac{\partial \bm{P}}{\partial \rho}$, contains $\left(\bm{I}-\rho\bm{W}\right)^{-1}$, and each iteration involves the inversion of an n-dimensional matrix. To circumvent the problem, \cite{klier2008clustering} linearized the estimator around the convenient starting point of standard Probit model, thus greatly simplified the gradient terms. In particular, they first estimate $\hat{\beta}_0$ with standard Probit model, compute the generalized error term $u_0=y_i-P\left(y_i=1\right)$ and the gradient terms $G_{\beta_i}$ and $G_{\rho_i}$. In the second step, they regress $G_{\beta_i}$ and $G_{\rho_i}$ on $Z$ to obtain the predictions $\hat{G}_{\beta_i}$ and $\hat{G}_{\rho_i}$. Then they regress $u_0+G_\beta'\hat{\beta}_0$ on $\hat{G}_{\beta_i}$ and $\hat{G}_{\rho_i}$ for the estimated values of $\beta$ and $\rho$.

\section{Empirical Results}

I focus on enterprises that operate in Shenzhen's computers, communications, and other electronic equipment manufacturing industry division in 2017, then estimate the spatial lagged Probit model for 2018, 2019, 2020, and 2021, at the industry group and industry class level respectively. In each model, I alter the specification of $W$ so that only spatial dependence among enterprises within the same classification category is allowed. For example, at the industry class level, I assume that the propensities to exit are correlated among computer parts manufacturers, but not between computer parts manufacturers and computer peripheral manufacturers. Enterprises within the same industry class produce homogeneous products. The industry group level also includes enterprises that operate in the upstream, downstream, or supporting industries of the business. Although it would be an interesting exploration, I have not estimated the model at the broader industry division and industry section level. This helps filter certain industry-specific characteristics, for example, while considering whether to exit the market, a computer parts manufacturer may not take into account the tendencies of nearby plastic product manufacturers.

\begin{table}[!htbp] \centering
  \caption{Neighborhood Effect on Enterprise Exits in Shenzhen's Manufacturing of Computers, Communications and other Electronic Equipment}
  \begin{tabular}{lcccc}
    \\[-1.8ex]\hline \hline \\[-1.8ex]
    & \multicolumn{1}{c}{2017-2018} & \multicolumn{1}{c}{2018-2019} & \multicolumn{1}{c}{2019-2020} & \multicolumn{1}{c}{2020-2021} \\
    \hline \\[-1.8ex]
    Industry Group & 0.77$^{***}$ & 1.00$^{***}$ & 1.65$^{***}$ & 2.28$^{***}$ \\
    & (0.25) & (0.38) & (0.43) & (0.43) \\
    Industry Class & $-$0.22 & $-$0.001 & 0.55 & 0.94 \\
    & (0.18) & (0.41) & (0.53) & (0.62) \\
    \hline \hline \\[-1.8ex]
    \textit{Note:}  & \multicolumn{4}{r}{$^{*}$p$<$0.1; $^{**}$p$<$0.05; $^{***}$p$<$0.01} \\
  \end{tabular}
\end{table}

Firstly, significant spillover effects are identified at the industry group level, but not at the industry class level. This implies that enterprises' propensities to exit may not be affected by their competitors, but are significantly influenced by enterprises that operate in their upstream or downstream business.

Secondly, at the industry group level, the spillover effect gradually becomes larger from 2018 to 2021. This could be explained by supply chain disruptions. The trade war leads to increased tariffs and restrictions on trade, which disrupts the supply chains. This could cause some foreign-invested enterprises to exit the market or reduce their operations, affecting the supply chains of other enterprises in the same industry. The COVID-19 pandemic further exacerbated supply chain disruptions due to lockdowns and restrictions on the movement of goods and people. These combined effects lead to an increased spillover effect, as more enterprises may consider exiting the market or scaling down their operations due to uncertainties and disruptions in the supply chain. In addition, both the trade war and the COVID-19 pandemic created significant uncertainty in the global market. This could lead to reduced foreign investment in China and increased risk aversion among foreign-invested enterprises. As a result, some enterprises might exit the market, leading to a larger spillover effect on the remaining enterprises in the same industry.

\begin{table}[!htbp] \centering
  \caption{Covariate Effects on Enterprise Exits in Shenzhen's Manufacturing of Computers, Communications and other Electronic Equipment}
  \setlength\tabcolsep{2pt}
  \begin{tabular}{lcccccccc}
    \\[-1.8ex]\hline \hline \\[-1.8ex]
    & \multicolumn{2}{c}{2017-18} & \multicolumn{2}{c}{2018-19} & \multicolumn{2}{c}{2019-20} & \multicolumn{2}{c}{2020-21} \\
    \\[-1.8ex] & Group & Class & Group & Class & Group & Class & Group & Class \\
    \hline \\[-1.8ex]
    Years of Operation & $-$0.13$^{***}$ & $-$0.13$^{***}$ & $-$0.08$^{**}$ & $-$0.08$^{**}$ & 0.03 & 0.04 & 0.02 & 0.04 \\
    & (0.03) & (0.03) & (0.04) & (0.04) & (0.04) & (0.04) & (0.04) & (0.04) \\
    Registered Capital & 0.21 & 0.21 & 0.09 & 0.10 & 0.23 & 0.28$^{*}$ & 0.21 & 0.24$^{*}$ \\
    & (0.24) & (0.24) & (0.23) & (0.23) & (0.16) & (0.15) & (0.13) & (0.12) \\
    Foreign Contributed & 0.12 & 0.11 & $-$0.29 & $-$0.30 & $-$0.25 & $-$0.29 & $-$0.06 & $-$0.11 \\
    & (0.26) & (0.26) & (0.25) & (0.25) & (0.25) & (0.25) & (0.25) & (0.25) \\
    Registration (Taiwan) & 0.19 & 0.21$^{*}$ & 0.34$^{**}$ & 0.33$^{**}$ & 0.27$^{*}$ & 0.27$^{*}$ & 0.19 & 0.19 \\
    & (0.12) & (0.12) & (0.14) & (0.14) & (0.14) & (0.14) & (0.14) & (0.14) \\
    Registration (U.S.) & $-$0.03 & $-$0.03 & $-$0.07 & $-$0.08 & $-$0.07 & $-$0.07 & 0.001 & $-$0.0004 \\
    & (0.19) & (0.19) & (0.20) & (0.20) & (0.21) & (0.21) & (0.21) & (0.22) \\
    Legal Form (Joint-venture) & 0.30$^{*}$ & 0.27$^{*}$ & 0.11 & 0.11 & $-$0.04 & $-$0.06 & 0.10 & 0.07 \\
    & (0.16) & (0.16) & (0.16) & (0.16) & (0.16) & (0.16) & (0.16) & (0.16) \\
    Is Tariffed & 0.25$^{**}$ & $-$0.36 & 0.05 & $-$0.22 & 0.15 & 0.31 & 0.21 & 0.56 \\
    & (0.11) & (0.32) & (0.15) & (0.43) & (0.14) & (0.40) & (0.13) & (0.39) \\
    Importer/Exporter & $-$0.22$^{***}$ & $-$0.23$^{***}$ & $-$0.10 & $-$0.10 & $-$0.04 & $-$0.04 & $-$0.03 & $-$0.04 \\
    & (0.06) & (0.06) & (0.07) & (0.07) & (0.07) & (0.07) & (0.07) & (0.07) \\
    \hline \hline \\[-1.8ex]
    \textit{Note:}  & \multicolumn{8}{r}{$^{*}$p$<$0.1; $^{**}$p$<$0.05; $^{***}$p$<$0.01} \\
  \end{tabular}
\end{table}

For the covariate effects, years of operation show a significant negative effect on market exits in 2017-18 and 2018-19. This suggests that younger firms are generally more likely to exit the market, but the effect may have lessened during 2019-20 and 2020-21. The size of the enterprises, as measured by registered capital, and the percentage of capital contributed by foreign investors, do not exhibit significant impacts on market exits across the years. Enterprises registered in Taiwan show a higher likelihood of exiting the market from 2017-2020 compared to those registered in Hong Kong. Enterprises registered in the U.S. have not shown any significant differences in market exit propensity. Joint ventures demonstrated a higher likelihood of exiting the market in 2018, but this effect is not significant in later years. Firms exposed to tariffs have a higher likelihood of market exits in 2018 at the industry group level, but no significant effects are observed in subsequent years or at the industry class level. Surprisingly, importers and exporters exhibit a negative and significant effect on market exits in 2018. This may be due to the time point that the trade war has not fully impacted the market. In later years, this effect was not significant, suggesting that the relationship between import/export status and market exits may have weakened over time. In addition, factors that are not captured in the model may also be at play. For instance, some importers and exporters might have been able to adapt to the changing trade environment by diversifying their markets, sourcing from alternative suppliers, or passing on the increased costs to customers. Government interventions, such as subsidies or other supportive policies, might have helped some affected enterprises weather the impact of the trade war, leading to an insignificant or mixed effect of importer and exporter status on market exits during this period.

\section{Conclusion}

This paper looks into how its neighboring firms may react when an electronic manufacturer in Shenzhen tends to exit in 2017-21. Using the spatial lagged Probit model, significant spillover effects are reported for enterprises that operate in the upstream or downstream industries, but not in the same industry. This indicates that the exit of an enterprise may change the dynamics of the supply chain, leading to positive spillover effects for firms in related industries. For example, the exit of a computer parts manufacturer could lead to an increase in demand for the parts of computer manufacturers. The model also captures the raising uncertainty and volatility brought by the pandemic and the U.S.-China trade war. These external shocks cause supply chain disruptions in Shenzhen's electronic industry, forcing enterprises to find alternative suppliers or change their production processes, potentially leading to more dependencies between firms in different industries. The age, size, registration area, and legal form also affect the firm's probability to exit.

Although this chapter provides the first empirical result of the spillover effects of Shenzhen electronic manufacturers' exit behaviors based on firm-level data, it has a few limitations. Firstly, I focus on 2017-2021, a time frame characterized by unique and dramatic shifts in global trade and public health. While the data provides a rich context for studying firm exit and spillover effects, it may limit the generalizability of these findings to other periods. Secondly, the sample is comprised of enterprises that have at least one foreign investor, and no local enterprises in Shenzhen are included. I alleviate the influence by focusing on the electronic manufacturing industry in Shenzhen, but the neighborhood effect may be better estimated by constructing a more comprehensive sample in the future or investigating other industries, such as the financial industry in Shanghai, and automobile manufacturers in Guangzhou. Thirdly, as is noted in the empirical result section, certain factors that influence enterprises' tendencies to exit and the spillover effects are not included in the model. For example, neighborhood effects may depend on events that happen locally, preferential policies, and macro factors, e.g. regional characteristics, industry structures, among others. Some policies may directly affect the spillover effect, e.g. China has implemented a number of policies to support industrial clustering in order to promote regional economic development and technological innovation, and improve the competitiveness of Chinese industries in the global market. Other effects are indirect, e.g. if the business environment of a city is friendly with macroeconomic stability, supporting policy and resources, then exits of enterprises in adjacent areas may generate smaller effects on one's exit decision. Some evidence is offered by previous research, e.g. \cite{fafchamps2013local} shows that the availability of local banks helps small and medium-sized enterprises expand and acquire investment \cite{fafchamps2013local}, and it may also aid enterprises' growth and reduce bankruptcy \cite{arcuri2020early}. From another angle, \cite{basile2017agglomeration} discovers that local industry variety could also alleviate the exit of enterprises. Omitting these important factors could result in bias in the estimation.

\newpage
\bibliographystyle{unsrt}  
\bibliography{references}

\end{document}